\documentclass[12pt]{article}
\usepackage[utf8]{inputenc}

\usepackage{makeidx}         
\usepackage{graphicx}        
\usepackage{multicol}        
\usepackage{multirow}
\usepackage[bottom]{footmisc}
\usepackage{geometry} 
\usepackage{amsmath, amssymb, amsthm}
\usepackage{psfrag,epsf}
\usepackage{enumerate}
\usepackage{url} 
\usepackage{hyperref}
\usepackage{booktabs}
\usepackage{xcolor}
\usepackage{natbib}
\usepackage[italian,english]{babel}
\usepackage{authblk}
\usepackage{changes}
\usepackage{subcaption}
\usepackage{bbm}
\usepackage{placeins}

\hypersetup{
			colorlinks=true,
			linkcolor=black,
            linktoc=page,
			anchorcolor=black,
			citecolor=blue,
			urlcolor=black,
}

\def \bs{\mathbf}

\newtheorem{proposition}{Proposition}

\def\0{\mbox{\bf{0}}}
\def\bs{\mathbf{s}}\def\be{\mathbf{e}}

\usepackage[toc,page]{appendix}

\def \be{\begin{equation}}
\def \ee{\end{equation}}
\def \ber{\begin{eqnarray}}
\def \eer{\end{eqnarray}}
\def \berr{\begin{eqnarray*}}
\def \eerr{\end{eqnarray*}}
\def \bqmatrix{\begin{bmatrix}}
\def \eqmatrix{\end{bmatrix}}

\def \be{\begin{equation}}
\def \ee{\end{equation}}
\def \ber{\begin{eqnarray}}
\def \eer{\end{eqnarray}}
\def \berr{\begin{eqnarray*}}
\def \eerr{\end{eqnarray*}}
\def \bamatrix{\begin{pmatrix}}
\def \eamatrix{\end{pmatrix}}
\def \bqmatrix{\begin{bmatrix}}
\def \eqmatrix{\end{bmatrix}}

\def \argmin{{\rm argmin }}

\def \argmin{{\rm argmin }}

\def \bs{\boldsymbol}

\def \qmo{``}

\def \qmcsp{'' }

\begin{document}
\title{\bf }
\author[1]{Beatrice Foroni}
\affil[1]{MEMOTEF Department, Sapienza University of Rome}
\author[2]{Luca Merlo\thanks{Corresponding author: luca.merlo@unier.it}}
\affil[2]{Department of Human Sciences, European University of Rome}
\author[1]{Lea Petrella}

\title{Expectile hidden Markov regression models for analyzing cryptocurrency returns}
%
%
\maketitle

\abstract{
In this paper we develop a linear expectile hidden Markov model for the analysis of cryptocurrency time series in a risk management framework. The methodology proposed allows to focus on extreme returns and describe their temporal evolution 
 by introducing in the model time-dependent coefficients evolving according to a latent discrete homogeneous Markov chain. As it is often used in the expectile literature, estimation of the model parameters is based on the asymmetric normal distribution. Maximum likelihood estimates are obtained via an Expectation-Maximization algorithm using efficient M-step update formulas for all parameters. 
 We evaluate the introduced method with both artificial data 
 under several experimental settings and real data investigating 
 the relationship between daily Bitcoin returns and major world market indices.}\\


\newpage

\section{Introduction}\label{sec1}
In the last ten years, investors have been increasingly attracted by the exploit of the cryptocurrency market, mostly because of its peculiar characteristics. Born merely as a peer-to-peer electronic cash system \citep{nakamoto2008bitcoin}, the $70$ billion increase in market capitalization (in particular Bitcoin during 2016-2017), enormous price jumps and levels of high volatility that were never seen before have made cryptos a new category of investment assets. Their unusual behavior makes them prone to some speculative bubbles that may in turn threaten the stability of financial markets \citep{cheah2015speculative, yarovaya2016intra}. Being crucial to address the level of integration between cryptocurrencies and traditional financial assets, many contributions have analyzed the relationship with equities mainly relying on well-known econometric techniques such as GARCH models \citep{katsiampa2019volatility, guesmi2019portfolio}, variance decomposition \citep{ji2018network, corbet2018exploring, yi2018volatility} and Granger causality test \citep{bouri2020volatility}. Part of the related literature has focused on extreme returns by using models capturing the tail behaviour, rather than inferring such occurrences from models based on conditional central tendency.
 For instance, \cite{kristjanpoller2020cryptocurrencies} and \cite{naeem2021asymmetric} employ a multifractal asymmetric analysis, indicating the presence of heterogeneity in the cross-relationship between most cryptocurrencies and equity ETFs and showing different behaviors between upward and downward trends. \cite{shahzad2022extreme} investigate tail-based connectedness  among major cryptocurrencies in extreme downward and upward market conditions using LASSO penalized quantile regressions, while \cite{zhang2021risk} apply a risk spillover approach based on generalized quantiles, showing the existence of a downside risk spillover between Bitcoin and traditional assets. 

In quantitative risk management, indeed, investigating the dynamic of extreme occurrences is of utmost importance for market participants and regulators. Among the different methods considered throughout the literature, 
quantile regression, introduced by \cite{koenker1978regression}, has represented a valid approach for modeling the entire distribution of returns while accounting for the well-known stylized facts, i.e., high kurtosis, skewness and serial correlation, that typically characterize financial assets. 
 In the financial literature, the quantile regression framework has been positively applied to estimate and forecast Value at Risk (VaR) and quantile-based risk measures (\citealt{engle2004CAViaR, white2015var, taylor2019forecasting, merlo2021}). 
\\
Several generalizations of the concept of 
 quantiles have also been introduced over the years. One important extension is provided by the expectile regression (\citealt{newey1987asymmetric}), which can be thought of as a generalization of the classical mean regression based on an asymmetric squared loss function. Similar to quantile regression, expectile regression allows to characterize the entire conditional distribution of a response variable, but possesses several advantages over the former. First, expectiles are more informative than quantiles since they rely on tail expectations whereas quantiles use only the information on whether an observation is below or above the predictor. Second, the squared loss is continuously differentiable which makes the estimators and their covariance matrix easier to compute using fast and efficient algorithms. 
 For these reasons, expectile models have been implemented in several fields, such as longitudinal data (\citealt{tzavidis2016longitudinal, alfo2017finite, barry2021new}), spatial analysis (\citealt{sobotka2012geoadditive, spiegel2020spatio}), life expectancy (\citealt{nigri2022relationship}), economics and finance (\citealt{taylor2008estimating, kim2016nonlinear, belliniDB17, bottone2021unified}). Especially in the context of risk management, expectiles have gained an important role as potential competitors to the VaR and the Expected Shortfall measures. 
 Indeed, they possess several interesting properties in terms of risk measures (see for instance \citealt{bellini2012isotonicity}, \citealt{Bellinietal14} and \citealt{ziegel2016coherence}), and are the only risk measure that is both coherent (\citealt{Artzner99}) and elicitable (\citealt{Lambertetal08}). 
 


 Moreover, when modeling financial time series, returns often exhibit a clustering behavior over time which cannot be captured by traditional homogeneous regression models. Risk managers and regulators are increasingly interested in determining whether, and how, their temporal evolution can be influenced by hidden variables, e.g., the state of the market, during tranquil and crisis periods. In this context, Hidden Markov Models (HMMs, see \citealt{macdonald1997hidden, zucchini2016hidden}) have been successfully employed in the analysis of time series data, with applications to asset allocation and stock returns as discussed in \cite{mergner2008time, de2013dynamic, nystrup2017long} and \cite{maruotti2019hidden}. Quantile regression methods have also been generalized to account for serial heterogeneity. 
 For example, 
 \cite{liu2016markov} consider a quantile autoregression in which the parameters are subject to regime shifts determined by the outcome of a latent, discrete-state Markov process, while \cite{adam2019model} propose a model-based clustering approach where groups are inferred from conditional quantiles; see also \cite{ye2016markov}, \cite{maruotti2021hidden} and\cite{merlo2022quantile} for other applications of regime-switching models to financial and environmental time series. In longitudinal data, \cite{farcomeni2012quantile} and \cite{marino2018mixed} introduce linear quantile regression models where time-dependent unobserved heterogeneity is described through dynamic coefficients that evolve 
 according to a homogeneous hidden Markov chain. Within a Bayesian framework, a quantile nonhomogeneous HMM for longitudinal data has been recently proposed by \cite{liu2021bayesian}. 
   %
 To the best of our knowledge, however, a HMM for estimating conditional expectiles has not yet been proposed in the literature. 
 \\
 Motivated by the advantages of expectiles and the versatility of HMMs, we develop a linear expectile hidden Markov regression model to analyze the tail relation between cryptocurrencies and traditional asset classes. 
 The method introduced allows to examine the entire conditional distribution of returns given the hidden state and potential covariates, where the dynamics of returns over time is described by state-specific regression coefficients which follow a latent discrete homogeneous Markov chain. Inference about model parameters is carried out in a Maximum Likelihood (ML) approach using an Expectation-Maximization (EM) algorithm based on the asymmetric normal distribution of \cite{waldmann2017bayesian} as working likelihood.
 From a risk management standpoint, the proposed methodology contributes to identify and control for potential inherent risks related to the participation in crypto exchanges to develop appropriate policies and risk assessment procedures.\\ 
The study period considered starts from September 2014 until October 2022, comprising numerous events that heavily impacted financial stability, as the Chinese stock market crash of 2015, the crypto currency bubble crisis in 2017-2018, the COVID-19 outbreak in 2020 and the Russian invasion of Ukraine at the beginning of 2022. Following \cite{corbet2018exploring}, we model Bitcoin daily returns as a function of major stock and global market indices, including Crude Oil, Standard $\&$ Poor's 500 (S$\&$P500), Gold COMEX daily closing prices and the Volatility Index (VIX). Our results show that Bitcoin returns exhibit a clear temporal clustering behavior in calm and turbulent periods, and they are strongly associated with traditional assets at low and high expectile levels. 
\\
In concluding, we also evaluate the performance of our approach in a simulation study, generating observations from a two-state HMM under two different sample sizes and two different distributions for the error terms. Additional simulation studies with different error distributions, a higher number of hidden states and a less persistent transition probability matrix are illustrated in the Supplementary Materials.\\


 The rest of the paper is organized as follows. Section \ref{sec:pre} briefly reviews the expectile regression. In Section \ref{sec:method} we specify the proposed model with the EM algorithm for estimating the model parameters and the computational aspects. In Section \ref{sec:sim}, we evaluate the performance of our proposal in a simulation study. 
 Section \ref{sec:emp} shows the empirical analysis and discusses the results obtained while Section \ref{sec:concl} concludes.

\section{Expectile regression}\label{sec:pre}
Expectile regression has been proposed by \cite{newey1987asymmetric} as a \qmo quantile-like\qmcsp generalization of standard mean regression based on asymmetric least-squares estimation. Similarly to quantile regression of \cite{koenker1978regression}, this is an alternative approach for characterizing the entire conditional distribution of a response variable where the quantile loss function is substituted with an asymmetric squared loss function. Formally, the expectile of order $\tau \in (0, 1)$ of a continuous response $Y$ given the $P$-dimensional vector of covariates $\bs X = \bs x$, is defined as the minimizer, $\mu_{\bs x} (\tau)$, of the following problem: 
\be\label{eq:squaredloss}
\mu_{\bs x} (\tau) = \underset{\mu \in \mathbb{R}}{\argmin} \, \mathbb{E}[\omega_\tau (Y - \mu_{\bs x} (\tau))],
\ee 
where $\omega_\tau(u) = u^2 \lvert \tau - \mathbb{I}(u < 0)\rvert$ is the asymmetric square loss and 
 $\mathbb{I}(\cdot)$ denotes the indicator function.\\
In a regression framework, for a given $\tau$, a linear expectile model is defined as $\mu_{\bs x} (\tau) = \bs x' \bs \beta (\tau)$, where $\bs \beta (\tau) \in \mathbb{R}^P$ is the regression parameter vector. If $\tau = \frac{1}{2}$, expectile regression reduces to the standard mean regression while for $\tau \neq \frac{1}{2}$ it allows to target the entire conditional distribution of the response given the covariates similarly to quantile regression. When we turn from quantiles to expectiles, the latter possess several advantages over the former. Particularly, 
 we gain uniqueness of the ML solutions which is, indeed, not granted in the quantile context. 
 From a computational standpoint, since the squared loss function $\omega_\tau(\cdot)$ is differentiable, the regression parameters $\bs \beta (\tau)$ can be estimated by efficient Iterative Reweighted Least Squares (IRLS), in contrast to algorithms used for fitting quantile regression models. Proofs of consistency, asymptotic normality and a robust estimator of the variance-covariance matrix of the regression coefficients for inference have been established in \cite{newey1987asymmetric}. These properties make the expectile regression versatile and computationally appealing from a statistical point of view. 
\\
 In a likelihood approach, \cite{gerlach2015bayesian} and \cite{waldmann2017bayesian} originally introduced the idea of 
 expectile regression by employing a likelihood function that is based on the Asymmetric Normal (AN) distribution. 
  The AN distribution can be thought of as a generalization of the normal distribution to allow for non-zero skewness, having the following density:
\be\label{eq:AND}
f_Y (y) = \frac{2\sqrt{\tau (1-\tau)}}{\sqrt{\pi \sigma^2} (\sqrt{\tau} + \sqrt{1-\tau})}\exp \left[ -\omega_\tau \left(\frac{y - \mu}{\sigma}\right)\right],
\ee
where $\mu \in \mathbb{R}$ is a location parameter corresponding to the $\tau$-th expectile of $Y$, $\sigma > 0$ is a scale parameter and $\tau \in (0,1)$ determines the asymmetry of the distribution. Particularly, when $\tau = \frac{1}{2}$ the density in \eqref{eq:AND} reduces to the well-known normal distribution, and $\mu$ and $\sigma$ coincide with its mean and standard deviation, respectively. 
 As discussed by \cite{waldmann2017bayesian}, the minimization of the asymmetric squared loss function in \eqref{eq:squaredloss} is equivalent, in terms of parameter estimates, to the maximization of the likelihood associated with the AN density.
\\ 
In the following section, we extend the expectile regression to the HMM setting by using the AN distribution as working likelihood. 


\section{Methodology}\label{sec:method}
In this section we describe the 
expectile hidden Markov regression model in order to take into account the temporal evolution of the time series under analysis.
We then show how inference about model parameters can be carried out in a ML approach using the AN distribution introduced in the previous section.

 Formally, let $\{ S_t \}_{t=1}^T$ be a latent, homogeneous, first-order Markov chain defined on the discrete state space $\{1,\dots,K\}$. Let $\pi_k = Pr(S_1=k)$ be the initial probability of state $k$, $k = 1,\dots,K$, and $\pi_{k\lvert j} = Pr(S_{t+1}=k \lvert S_t=j)$, with $\sum_{k=1}^K \pi_{k \lvert j} = 1$ and $\pi_{k\lvert j} \geq 0$, denote the transition probability between states $j$ and $k$, that is, the probability to visit state $k$ at time $t+1$ from state $j$ at time $t$, $j,k = 1,\dots,K$ and $t=1,\dots,T$. More concisely, we collect the initial and transition probabilities in the $K$-dimensional vector $\bs \pi$ and in the $K \times K$ matrix $\bs \Pi$, respectively.
 
To build the proposed model, let $Y_t$ denote a continuous observable response variable and $\bs X_t= (1, X_{t2},\dots, X_{tP})'$ be a vector of $P$ exogenous covariates, with the first element being the intercept, at time $t=1,\dots,T$. For a given expectile level $\tau \in (0,1)$, the proposed linear Expectile Hidden Markov Model (EHMM) is defined as follows:
\be\label{eq:model_reg}
Y_t = \sum_{k=1}^K \mathbbm{1}_{S_t = k} (\bs X'_t \bs\beta_k(\tau) + \epsilon_{tk}(\tau)),
\ee 
with $\bs\beta_k(\tau) = (\beta_{1k} (\tau), \dots, \beta_{Pk} (\tau))' \in \mathbb{R}^P$ being a state-specific coefficient vector that assumes one of the values $\{\bs\beta_1(\tau), \dots, \bs\beta_K(\tau)\}$ depending on the outcome of the unobservable Markov chain $S_t$ and where $\epsilon_{tk}(\tau)$ is the error term whose conditional $\tau$-th expectile is assumed to be zero. 
\\
Extending the approach of \cite{waldmann2017bayesian} to the HMM setting, we use the AN distribution to describe the conditional distribution of the response given covariates and the state occupied by the latent process at time $t$, whose probability density function is now given by
\be\label{eq:cond_density}
f_Y (y_t \lvert \bs X_t = \bs x_t, S_t = k) = \frac{2\sqrt{\tau (1-\tau)}}{\sqrt{\pi \sigma^2_k} (\sqrt{\tau} + \sqrt{1-\tau})}\exp \left[ -\omega_\tau \left(\frac{y_t - \mu_{tk}}{\sigma_k}\right)\right],
\ee
where the location parameter $\mu_{tk}$ is defined by the linear model $\mu_{tk} = \bs x'_t \bs \beta_k (\tau)$.

Following the work of \cite{gassiat2016inference}, to ensure sufficient conditions for model identifiability we state the following Proposition.
\begin{proposition} For any fixed expectile level $\tau$, let's assume that the number of hidden states K is known, that the matrix of transition probability between states $\bs{\Pi}$ has full rank and that the K's conditional distributions of the response given covariates and the state occupied by the latent process are independent. Then, the model in (\ref{eq:model_reg})-(\ref{eq:cond_density}) is identifiable from the distribution of three consecutive variables $Y_1,Y_2,Y_3$, up to label swapping of the hidden states. 
\end{proposition}

The proof of this result can be obtained by adapting Theorem 1 in \cite{gassiat2016inference}.

In the following section we use the AN distribution as a working likelihood for estimating the model parameters in a regression framework.

\subsection{Likelihood inference}\label{subsec:inference}
In this section we consider a ML approach to make inference on model parameters. As is common for HMMs, and for latent variable models in general, we develop an EM algorithm (\citealt{baum1970maximization}) to estimate the parameters of the method proposed based on the observed data. To ease the notation, unless specified otherwise, hereinafter we omit the expectile level $\tau$, yet all model parameters are allowed to depend on it.
\\
For a given number of hidden states $K$, the EM algorithm runs on the complete log-likelihood function of the model introduced, which is defined as	 
\be
\begin{aligned}\label{eq:completel}
\ell_c(\bs \theta_\tau) = \sum_{k=1}^K {\gamma_1}(k)\log\pi_k &+ \sum_{t=2}^T\sum_{k=1}^K \sum_{j=1}^K {\xi_t}(j,k) \log \pi_{k\lvert j} \\ &+ \sum_{t=1}^T\sum_{k=1}^K {\gamma_t}(k)\log f_Y(y_t \lvert \bs{x}_t, S_t = k),
\end{aligned}
\ee
where $\bs \theta_\tau = (\bs \beta_1, \dots, \bs \beta_K, \sigma_1, \dots, \sigma_K, \bs \pi, \bs \Pi)$ represents the vector of all model parameters, ${\gamma_t}(k)$ denotes a dummy variable equal to $1$ if the latent process is in state $k$ at occasion $t$ and 0 otherwise, and ${\xi_t}(j,k)$ is a dummy variable equal to $1$ if the process is in state $j$ in $t-1$ and in state $k$ at time $t$ and $0$ otherwise. 

To estimate $\bs \theta_\tau$, the algorithm iterates between two steps, the E- and M-steps, until convergence, as outlined below.

\subsubsection*{E-step:} 
In the E-step, at the generic $(h+1)$-th iteration, the unobservable indicator variables ${\gamma_t}(k)$ and ${\xi_t}(j,k)$ in \eqref{eq:completel} are replaced by their conditional expectations given the observed data and the current parameter estimates $\bs \theta_\tau^{(h)}$. To compute such quantities we require the calculation of the probability of being in state $k$ at time $t$ given the observed sequence 
\be\label{gamma}
\gamma_t^{(h)}(k) = P_{\bs \theta_\tau^{(h)}}(S_t = k \lvert y_1,\dots,y_T )
\ee
and the probability that at time $t-1$ the process is in state $j$ and then in state $k$ at time $t$, given the observed sequence
\be\label{xi} 
\xi_t^{(h)}(j,k) = P_{\bs \theta_\tau^{(h)}}(S_{t-1} = j, S_t = k \lvert y_1,\dots,y_T ).
\ee
The quantities in \eqref{gamma} and \eqref{xi} can be obtained using the Forward-Backward algorithm of \cite{welch2003hidden}. Then, we use these to calculate the conditional expectation of the complete log-likelihood function in \eqref{eq:completel} given the observed data and the current estimates:
\be
\begin{aligned}\label{eq:estep}
Q(\bs \theta_\tau\lvert \bs \theta_\tau^{(h)}) = \sum_{k=1}^K \gamma^{(h)}_1(k)\log\pi_k &+ \sum_{t=2}^T\sum_{k=1}^K \sum_{j=1}^K \xi^{(h)}_t(j,k) \log \pi_{k\lvert j} \\ &+ \sum_{t=1}^T\sum_{k=1}^K \gamma^{(h)}_t(k)\log f_Y(y_t \lvert \bs{x}_t, S_t = k).
\end{aligned}
\ee

\subsubsection*{M-step:} 
In the M-step we maximize $Q(\bs \theta_\tau \lvert \bs \theta_\tau^{(h)})$ in \eqref{eq:estep} with respect to $\bs \theta_\tau$ to obtain the update parameter estimates $\bs \theta_\tau^{(h+1)}$. The maximization of $Q(\bs \theta_\tau \lvert \bs \theta_\tau^{(h)})$ can be partitioned into orthogonal subproblems, where the updating formulas for the hidden Markov chain and state-dependent regression parameters are obtained independently maximizing each of these terms. Formally, the initial probabilities $\pi_k$ and transition probabilities $\pi_{k\lvert j}$ are updated using:
\be
\pi^{(h+1)}_k = \gamma^{(h)}_1(k), \quad k = 1,\dots,K
\ee
and
\be
\pi^{(h+1)}_{k\lvert j} = \frac{\sum_{t=2}^T \xi^{(h)}_t(j,k)}{\sum_{t=2}^T \sum_{k=1}^K \xi^{(h)}_t(j,k)}, \quad j,k = 1,\dots,K.
\ee
To update the regression coefficients, the first-order condition of \eqref{eq:estep} with respect to $\bs \beta_k$, $k = 1,\dots,K$, yields
\be
\frac{\partial Q(\bs \theta_\tau \lvert \bs \theta_\tau^{(h)})}{\partial \bs \beta_k} \propto \sum_{t=1}^T \gamma^{(h)}_t(k) \lvert \tau - \mathbb{I}(y_t < \bs x'_t \bs \beta_k)\lvert \bs x_t (y_t - \bs x'_t \bs \beta_k) = \bs 0_P,  
\ee
so the M-step update expression for $\bs \beta_k$ is
\be\label{eq:betaupdate}
\bs \beta^{(h+1)}_k = \Big( \sum_{t=1}^T \gamma^{(h)}_t(k) \lvert \tau - \mathbb{I}(y_t < \bs x'_t \bs \beta_k)\lvert  \bs x_t \bs x'_t \Big)^{-1} \Big( \sum_{t=1}^T \gamma^{(h)}_t(k) \lvert \tau - \mathbb{I}(y_t < \bs x'_t \bs \beta_k)\lvert  \bs x_t y_t \Big), 
\ee
which can be computed using IRLS for cross-sectional data with appropriate weights. Similarly, from the first-order condition of \eqref{eq:estep} with respect to the scale parameters we obtain the following M-step update formula for $\sigma^2_k$:
\be
\sigma^2_k{}^{(h+1)} = \frac{2}{\sum_{t=1}^T \gamma^{(h)}_t(k)} \sum_{t=1}^T \gamma^{(h)}_t(k) \lvert \tau - \mathbb{I}(y_t < \bs x'_t \bs \beta^{(h+1)}_k)\lvert  (y_t - \bs x'_t \bs \beta^{(h+1)}_k)^2.
\ee
\\
The E- and M- steps are alternated until convergence, that is when the observed likelihood between two consecutive iterations is smaller than a predetermined threshold. In this paper, we set this threshold criterion equal to $10^{-4}$. 
\\
Following \cite{maruotti2021hidden} and \cite{merlo2022quantile}, for fixed $\tau$ and $K$ we initialize the EM algorithm by providing the initial states partition, $\{ S_t^{(0)} \}_{t=1}^T$, according to a Multinomial distribution with probabilities $1/K$. From the generated partition, the elements of $\bs \Pi^{(0)}$ are computed as proportions of transition, while we obtain $\bs \beta^{(0)}_k$ and $\sigma^{(0)}_k$ by fitting mean regressions on the observations within state $k$. To deal with the possibility of multiple roots of the likelihood equation and better explore the parameter space, 
 we fit the proposed EHMM using a multiple random starts strategy with different starting partitions and retain the solution corresponding to the maximum likelihood value.
\\
Once we computed the ML estimate of the model parameters, to estimate the standard errors we employ the parametric bootstrap scheme of \cite{visser2000confidence}. In practice, we refit the model to $R$ bootstrap samples and approximate the standard error of each model parameter with the corresponding standard deviation of the bootstrap estimates. 

\section{Simulation study}\label{sec:sim}
In this section we conduct a simulation study to validate the performance of our method under different scenarios in terms of:  (i) recovering the true values of the parameters; (ii) assessing the classification behavior of the proposed model; (iii) evaluating the capability of penalized likelihood criteria in selecting the optimal number of hidden states $K$. 
We analyze three sample sizes ($T=100,\ T=500,\ T=1000$) and two distributions for the error term. 
 For each scenario we conduct $500$ Monte Carlo simulations. We draw observations from a two state HMM ($K=2$) using the following data generating process:
\be\label{eq:sim}
Y_t = \sum_{k=1}^2 \mathbbm{1}_{S_t = k} (\bs X'_t \bs\beta_k(\tau) + \epsilon_{tk}(\tau))
\ee

with $\bs X_t= (1, X_{t2})'$, where $X_{t2} \sim \mathcal{N}(0,1)$, and with $\bs\beta_1(\tau) = (-1, 2)'$ and $\bs\beta_2(\tau) = (1, -2)'$.
We consider two distributions for the error terms in \eqref{eq:sim}. In the first scenario, $\epsilon_{tk}$ is generated from a Gaussian distribution with standard deviation 1, for $k=1,2$. In the second one, $\epsilon_{tk}$ is generated from a skew-$t$ distribution with 5 degrees of freedom and asymmetry parameter 2, for $k=1,2$. 
Finally, the matrix of transition probabilities is set equal to $\bs \Pi = \bigl( \begin{smallmatrix} 0.8 & 0.2\\ 0.2 & 0.8\end{smallmatrix}\bigr)$,
 while the vector of initial probabilities is equal to $\bs \pi = (1,0)$.
 Additional simulation studies with different error distributions, a higher number of hidden states and a less persistent transition probability matrix are illustrated in the Supplementary Materials.

\begin{figure}[!h]
\centering
\includegraphics[width=.65\linewidth]{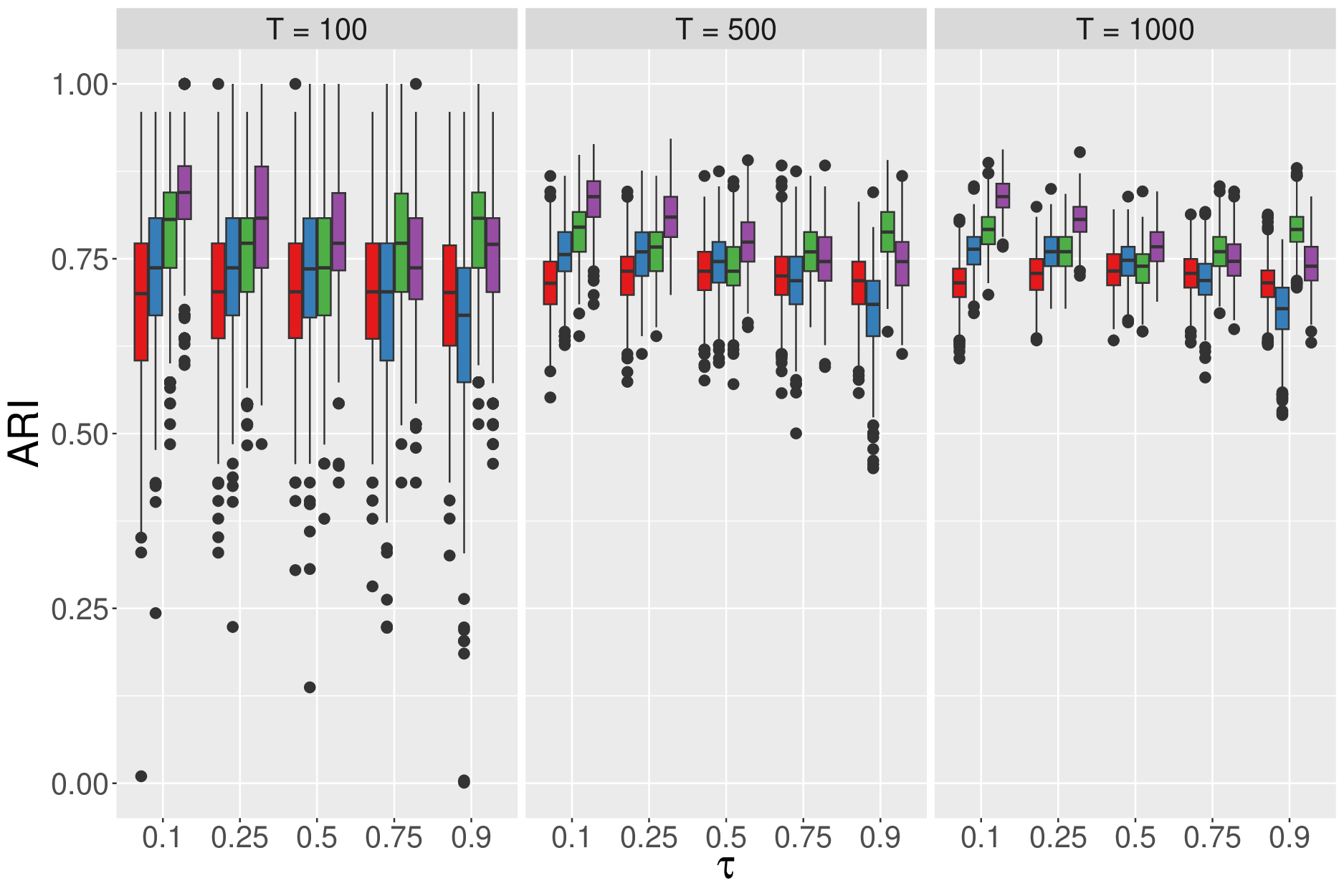}
\caption{From left to right, box-plots of ARI for the classification obtained according to the posterior probabilities for Gaussian (red) and skew-$t$ (blue) distributed errors with $T=100$, $T=500$ and $T=1000$. 
In green and purple are represented the box-plots of ARI related to the partitions obtained by considering the true model parameters, with Gaussian and skew-$t$ distribution respectively.}
\label{fig:ari_persmat}
\end{figure}

In order to assess the validity of the model we fit the proposed EHMM at five expectile levels, i.e., $\tau = \{0.10, 0.25, 0.50, 0.75, 0.90\}$, and compute the bias and standard errors associated to the state-specific coefficients, averaged over the Monte Carlo replications, for each combination of sample size and error distribution. Tables \ref{tab:gaus_perst} and \ref{tab:skewt_perst} report the simulation outputs for the Gaussian and skew-$t$ distributions, respectively. 
 As can be observed, as regards Gaussian distributed errors, the precision of the estimates is higher at the center of the distribution rather than on the tails, mainly due to the reduced number of observations at extreme expectile levels, but the bias always remains under control. Evidently, in Table \ref{tab:skewt_perst} a higher standard deviation shows up for the skew-$t$ distribution due to the asymmetry and heavier tails than the Gaussian density, but both the bias and the standard deviation tend to decrease as the sample size increases, with some exception, probably related to Monte Carlo variability. 
 Concerning the hidden process, given the true values of the transition probabilities in $\bs \Pi$, we see that the coefficients corresponding to the first state are estimated with lower precision because fewer transitions occur from one state to the other, as expected.\\
 To evaluate the ability in recovering the true states partition we consider 
 the Adjusted Rand Index (ARI) of \cite{hubert1985comparing}. The state partition provided by the fitted models is obtained by taking the maximum, $\underset{k}{\max} \, \gamma_{t} (k)$, posteriori probability for every $t = 1,\dots,T$, and report the box-plots of ARI for the classification obtained according to the posterior probabilities in Figure \ref{fig:ari_persmat} for the four settings considered. As a benchmark, we also report box-plots of ARI related to the partitions obtained by considering the true model parameters. 
 Firstly, we observe that the accuracy of estimating the true state partition is significantly influenced by distribution error, across all five expectile levels. 
 The value of the asymmetry parameter and the tail-heaviness determine better results for the skew-$t$ in the left tail of the distribution, while exhibiting a comparatively inferior performance in the right tail with respect to the Gaussian distribution.
 Secondly, the goodness of the clustering obtained partially depends on the specific expectile level as regards the Gaussian distribution, being the values slightly higher at the mean ($\tau = 0.50$) than at the tails. Finally, when increasing the sample size from $T=100$ to $T=500$ and $T=1000$, results clearly improve reporting a lower variability for both error distributions. Overall, the proposed EHMM is able to recover the true values of the parameters and the true state partition highly satisfactory in all the cases examined. 
  The last goal of this simulation exercise is to assess the performance of three widely employed penalized likelihood criteria for selecting the true number of hidden states $K$, namely the AIC (\citealt{akaike1998information}), the BIC (\citealt{schwarz1978estimating}) and the ICL (\citealt{biernacki2000assessing}). Following the work of \cite{merlo2022quantile}, we use the same generating data process in (\ref{eq:sim}), drawing observations from a two state HMM ($K = 2$) with $T = 2000$. We fit the EHMM with $K = 1,2,3,4$ in order to select the best $K$ associated to the lowest penalized likelihood criteria over 300 Monte Carlo replicates. Table \ref{tab:criteria_eval} reports the percentage frequency distributions of the selected $K$ for each of the three criteria at three expectile levels, i.e. $\tau = \{ 0.10, 0.50, 0.90\} $, for Gaussian and skew-$t$ distributed errors (we do not show results for $K = 1$ since it is never selected by any criteria). We can observe that the BIC and ICL work well at $\tau = 0.50$ and for Gaussian distributed errors but, as we move towards the tails of the distribution of the response variable, ICL always outperforms the other criteria. 
  These results suggest that, among the criteria considered, the ICL 
  does well in terms of correctly identifying
  the number of latent states across all considered scenarios,
  capturing serial heterogeneity in the data in a more parsimonious manner.
\begin{table}[h]
  \centering
  \scalebox{0.7}{
\begin{tabular}{lrrrrrrrrrr}
$\tau$ & 0.10  &       & 0.25  &       & 0.50  &       & 0.75  &       & 0.90  &  \\
\midrule
      & Bias  & Std.Err & Bias  & Std.Err & Bias  & Std.Err & Bias  & Std.Err & Bias  & Std.Err \\
\midrule
Panel A: T=100 &       &       &       &       &       &       &       &       &       &  \\
State 1 &       &       &       &       &       &       &       &       &       &  \\
$\beta_{1,1}$ = -1 & 0.00622 & 0.33093 & 0.00955 & 0.12356 & -0.00565 & 0.11848 & -0.02963 & 0.13449 & -0.06976 & 0.17636 \\
$\beta_{2,1}$ = 2 & -0.01692 & 0.26157 & 0.00081 & 0.16778 & 0.00573 & 0.16098 & 0.01463 & 0.16816 & 0.03025 & 0.19427 \\
State 2 &       &       &       &       &       &       &       &       &       &  \\
$\beta_{1,2}$ = 1 & 0.08047 & 0.24807 & 0.04623 & 0.13363 & 0.01166 & 0.12077 & -0.01745 & 0.13365 & -0.04202 & 0.18725 \\
$\beta_{2,2}$ = -2 & 0.00566 & 0.30633 & -0.00625 & 0.17357 & -0.00682 & 0.16814 & -0.00495 & 0.18425 & -0.00012 & 0.22838 \\
\midrule
      &       &       &       &       &       &       &       &       &       &  \\
Panel A: T=500 &       &       &       &       &       &       &       &       &       &  \\
State 1 &       &       &       &       &       &       &       &       &       &  \\
$\beta_{1,1}$ = -1 & 0.02575 & 0.07014 & 0.01263 & 0.05784 & 0.00044 & 0.05497 & -0.0171 & 0.06197 & -0.04871 & 0.0821 \\
$\beta_{2,1}$ = 2 & -0.01612 & 0.07496 & -0.00342 & 0.06725 & 0.00168 & 0.06623 & 0.00116 & 0.07122 & -0.00477 & 0.08286 \\
State 2 &       &       &       &       &       &       &       &       &       &  \\
$\beta_{1,2}$ = 1 & 0.05135 & 0.0778 & 0.01864 & 0.05887 & -0.00159 & 0.05347 & -0.0171 & 0.05688 & -0.03274 & 0.07033 \\
$\beta_{2,2}$ = -2 & 0.00684 & 0.08549 & 0.00288 & 0.07153 & 0.00269 & 0.0671 & 0.00642 & 0.07113 & 0.01655 & 0.08229 \\
\midrule
      &       &       &       &       &       &       &       &       &       &  \\
Panel B: T = 1000 &       &       &       &       &       &       &       &       &       &  \\
State 1 &       &       &       &       &       &       &       &       &       &  \\
$\beta_{1,1}$ = -1 & 0.02282 & 0.04637 & 0.01032 & 0.03886 & -0.00241 & 0.03703 & -0.02037 & 0.04082 & -0.0507 & 0.05365 \\
$\beta_{2,1}$ = 2 & -0.01544 & 0.05691 & -0.00562 & 0.04951 & -0.00272 & 0.04693 & -0.00338 & 0.04991 & -0.00793 & 0.0589 \\
State 2 &       &       &       &       &       &       &       &       &       &  \\
$\beta_{1,2}$ = 1 & 0.04491 & 0.05364 & 0.01622 & 0.04135 & -0.00129 & 0.03711 & -0.01424 & 0.03884 & -0.02716 & 0.04679 \\
$\beta_{2,2}$ = -2 & 0.00666 & 0.06116 & 0.00151 & 0.05149 & 0.00003 & 0.0471 & 0.00278 & 0.04753 & 0.01253 & 0.05389 \\
\bottomrule
\end{tabular}%
}
    \caption{Bias and standard error values of the state-regression parameter estimates with Gaussian distributed errors for $T=100$ (Panel A), $T=500$ (Panel B) and $T=1000$ (Panel C) for each expectile level considered.}
  \label{tab:gaus_perst}%
\end{table}%

\begin{table}[h]
  \centering
  \scalebox{0.7}{
\begin{tabular}{lrrrrrrrrrr}
$\tau$ & 0.10  &       & 0.25  &       & 0.50  &       & 0.75  &       & 0.90  &  \\
\midrule
      & Bias  & Std.Err & Bias  & Std.Err & Bias  & Std.Err & Bias  & Std.Err & Bias  & Std.Err \\
\midrule
Panel A: T=100 &       &       &       &       &       &       &       &       &       &  \\
State 1 &       &       &       &       &       &       &       &       &       &  \\
$\beta_{1,1}$ = -1 & 0.00815 & 0.1085 & -0.01061 & 0.10852 & -0.03357 & 0.18523 & -0.08009 & 0.32874 & -0.18035 & 0.48869 \\
$\beta_{2,1}$ = 2 & -0.00918 & 0.17167 & -0.00311 & 0.12744 & -0.00416 & 0.18517 & -0.03466 & 0.47901 & -0.11981 & 0.82864 \\
State 2 &       &       &       &       &       &       &       &       &       &  \\
$\beta_{1,2}$ = 1 & 0.01825 & 0.15049 & -0.00137 & 0.11403 & -0.0142 & 0.16168 & 0.04334 & 1.09454 & 0.11931 & 1.47897 \\
$\beta_{2,2}$ = -2 & 0.00479 & 0.23544 & 0.00591 & 0.14285 & -0.00152 & 0.4782 & 0.11121 & 0.96889 & 0.17729 & 0.64891 \\
\midrule
      &       &       &       &       &       &       &       &       &       &  \\
Panel A: T=500 &       &       &       &       &       &       &       &       &       &  \\
State 1 &       &       &       &       &       &       &       &       &       &  \\
$\beta_{1,1}$ = -1 & 0.00483 & 0.04542 & -0.01193 & 0.04596 & -0.03978 & 0.06474 & -0.09664 & 0.11736 & -0.22504 & 0.25398 \\
$\beta_{2,1}$ = 2 & -0.00604 & 0.05776 & -0.00186 & 0.05428 & 0.00372 & 0.0682 & 0.02162 & 0.11595 & 0.093 & 0.27504 \\
State 2 &       &       &       &       &       &       &       &       &       &  \\
$\beta_{1,2}$ = 1 & 0.00974 & 0.04828 & -0.00616 & 0.04369 & -0.01681 & 0.05402 & -0.02279 & 0.0903 & 0.0038 & 0.19797 \\
$\beta_{2,2}$ = -2 & 0.00925 & 0.05833 & 0.00865 & 0.05523 & 0.01502 & 0.06355 & 0.03747 & 0.08278 & 0.06971 & 0.22292 \\
\midrule
      &       &       &       &       &       &       &       &       &       &  \\
Panel B: T = 1000 &       &       &       &       &       &       &       &       &       &  \\
State 1 &       &       &       &       &       &       &       &       &       &  \\
$\beta_{1,1}$ = -1 & 0.00215 & 0.03171 & -0.01439 & 0.0314 & -0.04228 & 0.04483 & -0.10444 & 0.0845 & -0.2491 & 0.18266 \\
$\beta_{2,1}$ = 2 & -0.00609 & 0.03989 & -0.0021 & 0.03919 & 0.00261 & 0.04837 & 0.02072 & 0.08124 & 0.09536 & 0.19691 \\
State 2 &       &       &       &       &       &       &       &       &       &  \\
$\beta_{1,2}$ = 1 & 0.0108 & 0.03395 & -0.00527 & 0.0308 & -0.01522 & 0.03795 & -0.01847 & 0.06424 & 0.01649 & 0.14547 \\
$\beta_{2,2}$ = -2 & 0.00912 & 0.04106 & 0.00717 & 0.03911 & 0.01368 & 0.04379 & 0.03687 & 0.05688 & 0.0817 & 0.07759 \\
\bottomrule
\end{tabular}%
 }
    \caption{Bias and standard error values of the state-regression parameter estimates with skew-$t$ distributed errors for $T=100$ (Panel A), $T=500$ (Panel B) and $T=1000$ (Panel C) for each expectile level considered.}
  \label{tab:skewt_perst}%
\end{table}%

\begin{table}[htbp]
  \centering
  \scalebox{0.7}{
\begin{tabular}{rrrrrrrrrr}
\multicolumn{1}{l}{\boldmath{}\textbf{$\tau$}\unboldmath{}} &       & 0.10  &       &       & 0.50  &       &       & 0.90  &  \\
\midrule
      & AIC   & BIC   & ICL   & AIC   & BIC   & ICL   & AIC   & BIC   & ICL \\
Panel A: Gaussian errors &       &       &       &       &       &       &       &       &  \\
K = 2 & 0     & 0     & 100   & 57    & 91    & 95    & 0     & 0     & 100 \\
K = 3 & 0     & 27    & 0     & 31    & 4     & 5     & 0     & 25    & 0 \\
K = 4 & 100   & 73    & 0     & 12    & 5     & 0     & 100   & 75    & 0 \\
      &       &       &       &       &       &       &       &       &  \\
Panel B: skew-$t$ errors &       &       &       &       &       &       &       &       &  \\
K = 2 & 0     & 0     & 57    & 0     & 0     & 92    & 0     & 0     & 52 \\
K = 3 & 0     & 0     & 42    & 0     & 0     & 8     & 0     & 0     & 47 \\
K = 4 & 100   & 100   & 1     & 100   & 100   & 0     & 100   & 100   & 1 \\
\bottomrule
\end{tabular}%

}%
      \caption{Percentage frequency distribution of the selected number of hidden states $K$ under Gaussian and skew-$t$ errors over 300 replications.}
  \label{tab:criteria_eval}%
\end{table}%


\FloatBarrier

\section{Empirical application}\label{sec:emp}
In this section we apply the methodology proposed to analyze the Bitcoin daily returns as a function of global leading financial indices. Over the last decade, cryptocurrencies and in particular the Bitcoin market played a leading role, attracting attentions of researchers and investors. Their peculiar characteristics, such their extreme price volatility, driven by market speculation and technology applications, often lead to price bubbles, euphoria and market instability. 
 In order to address these periods of upheaval, it is crucial to understand the association between Bitcoin and globally relevant market indices in such circumstances of financial turmoil. Consistently with \cite{corbet2018exploring}, here we consider the Bitcoin, Crude Oil, Standard $\&$ Poor's 500 (S$\&$P500), Gold COMEX daily closing prices and the Volatility Index (VIX) from September 2014 to October 2022. 
 All series are expressed in US dollars and have been downloaded from the Yahoo finance database. Daily returns with continuous compounding are calculated 
 taking the difference of the logarithms between closing prices in consecutive trading days and then multiplied by 100, i.e., $r_t = 100 \cdot \log(P_t/P_{t-1})$, where $P_t$ is the closing price on day $t$, for a total of $T = 2025$ observations. 
\\
The considered timespan is marked by numerous crises that may have impacted cross-market integration patterns, including the Chinese stock market crash of 2015, the cryptocurrency bubble crisis in 2017-2018, the COVID-19 pandemic and the Russian invasion of Ukraine at the beginning of 2022, which have caused unprecedented levels of uncertainty and risk.
 In Table \ref{tab:summary} we report the list of examined variables and the summary statistics for the whole sample. First thing to notice is that Bitcoin is generally much more volatile than the other assets, having the highest standard deviation. The Bitcoin returns also show very high negative skewness and very high kurtosis, as well as S$\&$P500. The highest level of kurtosis is reported by Crude Oil, probably determined by the prices' fluctuations after the COVID-19 outbreak. On the other side, the large positive skewness of VIX indicates longer and fatter tails on the right side of the distribution, highlighting the well-known inverse relationship with the S$\&$P500.
   In concluding, the Augmented Dickey-Fuller (ADF) test \cite{dickey1979distribution} shows that all daily returns are stationary at the $1\%$ level of significance. Following these considerations, and motivated by the reforms considered by markets authorities to protect investors and preserve stability in response to cryptocurrencies' downturns, 
 the proposed EHMM can provide insights into the temporal evolution of Bitcoin returns and describe how this is affected by rapid changes in markets volatility.

\begin{table}[h]
  \centering
  \scalebox{0.8}{
\begin{tabular}{lrrrrrrrr}
      & \multicolumn{1}{c}{Minimum} & \multicolumn{1}{c}{Mean} & \multicolumn{1}{c}{Maximum} & \multicolumn{1}{c}{Std.Err.} & \multicolumn{1}{c}{Skewness} & \multicolumn{1}{c}{Kurtosis} & \multicolumn{1}{c}{Jarque-Bera test} & \multicolumn{1}{c}{ADF test} \\
\toprule
Bitcoin & -46.4730 & 0.1859 & 22.5119 & 4.6165 & -0.6817 & 8.7172 & \textbf{6568.469} & \textbf{-11.117} \\
Crude Oil & -28.2206 & 0.0280 & 31.9634 & 3.1077 & 0.0942 & 21.1219 & \textbf{37645.730} & \textbf{-10.324} \\
S\&P500 & -12.7652 & 0.0316 & 8.9683 & 1.1716 & -0.9033 & 16.3473 & \textbf{22823.380} & \textbf{-12.390} \\
Gold  & -5.1069 & 0.0152 & 5.7775 & 0.9358 & -0.0698 & 4.1741 & \textbf{1471.712} & \textbf{-12.292} \\
VIX   & -29.9831 & 0.0379 & 76.8245 & 8.3414 & 1.2683 & 6.6648 & \textbf{4290.737} & \textbf{-14.209} \\
\bottomrule
\end{tabular}%
    }
    \caption{Descriptive statistics for the whole sample. The Jarque-Bera test and the ADF test statistics are displayed in boldface when the null hypothesis is rejected at the $1\%$ significance level.}
  \label{tab:summary}%
\end{table}%
 
To this end, we consider the following linear EHMM:
\begin{multline}
	\mu_{tk}^{Bitcoin} = \beta_{1k}(\tau) + \beta_{2k}(\tau)r_{t}^{Crude \, \, Oil} + \beta_{3k}(\tau)r_{t}^{S \& P500} 
	+ \beta_{4k}(\tau)r_{t}^{Gold} + \beta_{5k}(\tau)r_{t}^{VIX},
\end{multline}
with $\mu_{tk}^{Bitcoin}$ corresponding to the $\tau$-th conditional expectile of Bitcoin return at time $t$ in state $k$, while $r_{t}^{Crude \, \, Oil}$ denotes the return of the same date for Crude Oil, and similarly for the other indices.
\\
As a first step of the empirical analysis, in order to select the number of latent states we fit the proposed EHMM for different values of $K$ varying from 2 to 5.
To model large negative and positive returns, we focus on three expectile levels $\tau = \{ 0.10, 0.50, 0.90 \}$.
To compare models with differing number of states, Table \ref{tab:aic_bic_icl} reports three widely employed penalized likelihood selection criteria for $K$, namely the AIC (\citealt{akaike1998information}), the BIC (\citealt{schwarz1978estimating}) and the ICL (\citealt{biernacki2000assessing}). As one can see, the AIC selects 5, or more, states, while BIC chooses $K = 4$ for all three expectile levels. This should not be surprising since the AIC tends to overestimate the true number of hidden states. On the contrary, ICL favors a more parsimonious choice as $K = 2$ is always considered to be the optimal number of states at $\tau = 0.10$, $\tau = 0.50$ and $\tau = 0.90$. For these reasons, and in order to clearly identify high and low volatility market conditions we thus consider the proposed EHMM with $K=2$ states for all three $\tau$ levels.\\
 
\begin{figure}[!h]
\centering
\includegraphics[width=.7\linewidth]{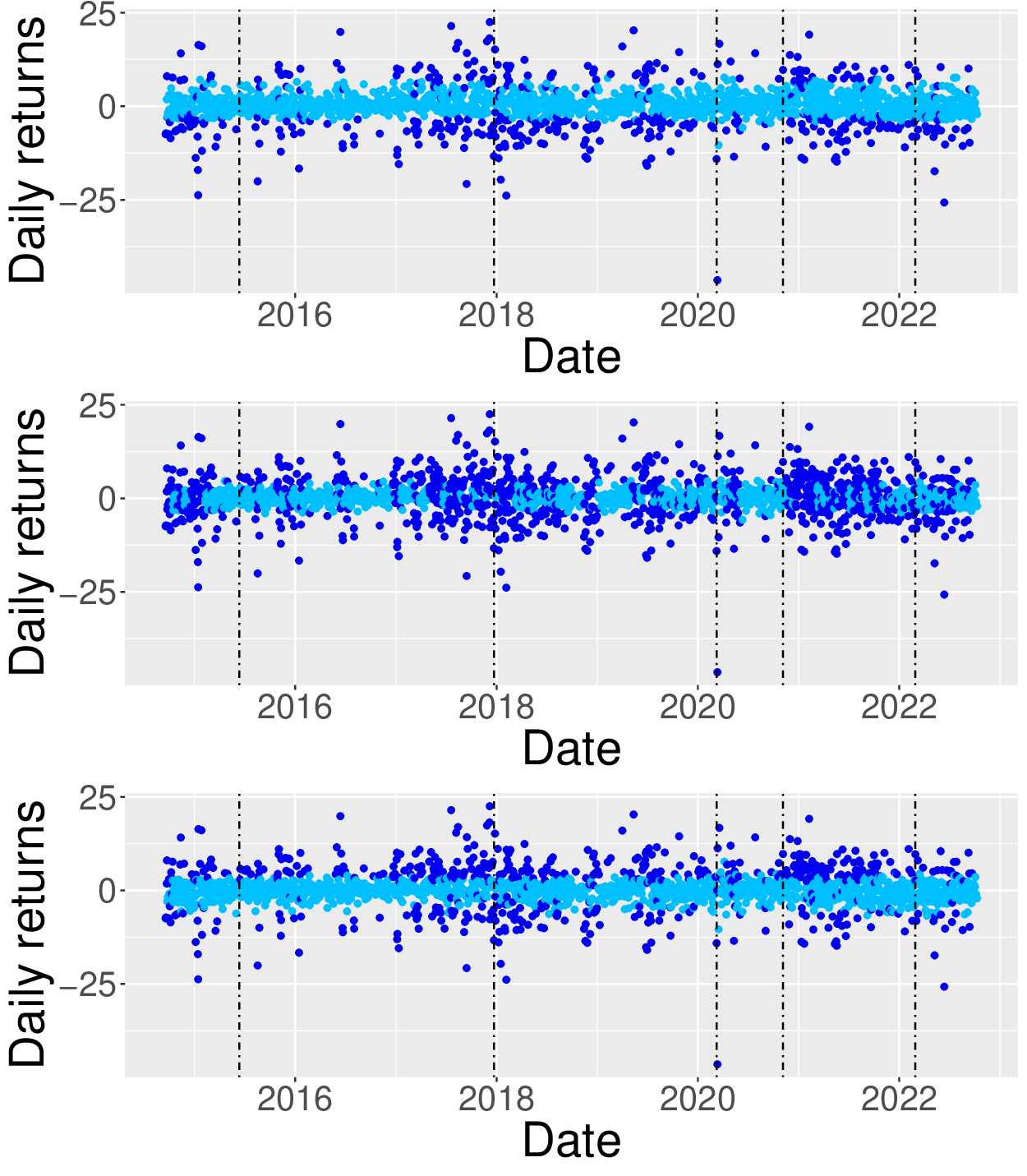}
\caption{From top to bottom, Bitcoin returns series classified according to the estimated posterior probability of class membership at $\tau = 0.10$, $\tau = 0.50$ and $\tau = 0.90$. Vertical dashed lines indicate globally relevant events in the financial markets that occurred in 2015,06; 2017,12; 2020,03; 2020,11; and 2022,02.}
\label{fig:cl_MAP}
\end{figure}

  For the selected models, we report the clustering results in Figure \ref{fig:cl_MAP} at the investigated expectile levels for Bitcoin daily returns, colored according to the estimated posterior probability of class membership, $\underset{k}{\max} \, \gamma_{t} (k)$, with the vertical dashed lines representing globally relevant events such as the Chinese stock market crash in 2015, the cryptocurrencies crash at the beginning of 2018, the COVID-19 market crash in March 2020, Biden's election at the USA presidency in November 2020 and the Russian invasion of Ukraine at the beginning of 2022. Here we clearly see that the latent components can be associated to specific market regimes characterized by low and high volatility periods. Specifically, light-blue points (State 1) tend to identify low returns, while dark-blue ones (State 2) correspond to periods of extreme positive and negative returns. 
When targeting the conditional mean ($\tau = 0.50$), as expected the level of separation among high and low volatility periods becomes less clear. 
 However, as the focus of our work lies especially at the tails of the returns distribution ($\tau = 0.10$ and $\tau = 0.90$), the classification obtained with $K=2$ is more clear and it allows us to distinguish low and high volatility periods.
 As a robustness check, we also used the Viterbi algorithm for the estimation of clustering partition (see Figure \ref{fig:cl_viterbi} in the Appendix). By comparing the results with those obtained with the MAP rule in Figure \ref{fig:cl_MAP}, the two assignment rules give the same results more than $90\%$ of the time.
 
Moving on to the state-specific model parameters, Table \ref{tab:parest} shows the parameter estimates along with the standard errors (in brackets) computed by using the parametric bootstrap technique illustrated in Section \ref{subsec:inference} over $R = 1000$ resamples. First, consistently with the quantile regression literature, the intercepts are increasing with $\tau$, with State 1 having lower values than State 2 for all $\tau$s. Second, it is interesting to observe that in the not-at-risk state (State 1) the S$\&$P500, Gold and the VIX index positively influence extreme left-tail ($\tau = 0.10$) movements of Bitcoin returns, while only S$\&$P500 and Gold significantly influence the right-tail ($\tau = 0.90$) expectiles of the cryptocurrency, exposing a connection during high volatility periods between traditional financial markets and Bitcoin both for negative and positive returns. At $\tau = 0.50$, instead, Bitcoin can be considered as a weak hedge during high volatility periods since it is not statistically associated with all the assets considered \citep{bouri2017bitcoin, bouri2017hedge}. In the at-risk state (State 2) we observe a positive influence of the S$\&$P500 and Gold across the conditional distribution of returns. Also, one can see that Crude Oil is negatively associated with the crypto returns at the $10$-th expectile. This finding is in line with \cite{bouri2020cryptocurrencies} but it is contrary to the works of \cite{dyhrberg2016hedging} and \cite{corbet2018exploring}, which may be due to the events and crises occurred in the last years.\\ 
 Finally, the estimated state-dependent scale parameter $\sigma_1$ reflects more stable periods for the first state, meanwhile $\sigma_2$ contemplates rapid (positive and negative) peak and burst returns for the second state, confirming the graphical analysis conducted in Figure \ref{fig:cl_MAP}.

\begin{table}[htbp]
  \centering
  \scalebox{0.8}{
\begin{tabular}{lrrr}
   & \multicolumn{1}{c}{$\tau = 0.10$} & \multicolumn{1}{c}{$\tau = 0.50$} & \multicolumn{1}{c}{$\tau = 0.90$} \\
\toprule
AIC \\
    $K=2$   & 11347.4122 & 11231.4286 & 11389.7954 \\
    $K=3$   & 11210.6074 & 11126.1327 & 11204.4727 \\
    $K=4$   & 11109.3930 & 11051.9556 & 11115.4655 \\
    $K=5$   & \textbf{11055.7980} & \textbf{11014.8696} & \textbf{11079.7912} \\
    \midrule
BIC \\
    $K=2$   & 11431.6121 & 11315.6285 & 11473.9953 \\
    $K=3$   & 11356.5539 & 11272.0791 & 11350.4192 \\
    $K=4$   & \textbf{11328.3127} & \textbf{11270.8753} & \textbf{11334.3852} \\
    $K=5$   & 11358.9175 & 11317.9892 & 11382.9108 \\
    \midrule
ICL \\
    $K=2$   & \textbf{12784.4362} & \textbf{12787.4471} & \textbf{12926.5182} \\
    $K=3$   & 13490.5599 & 13406.8221 & 13616.8609 \\
    $K=4$   & 13880.2346 & 13487.8461 & 13759.7332 \\
    $K=5$   & 14111.8066 & 13511.4108 & 13908.7362 \\ 
\bottomrule
\end{tabular}%
}
     \caption{AIC, BIC and ICL values with varying number of states for the investigated expectile levels. Bold font highlights the best values for the considered criteria (lower-is-better).}
     \label{tab:aic_bic_icl}%
\end{table}%

\begin{table}[htbp]
  \centering
  \resizebox{1.0\columnwidth}{!}{%
\begin{tabular}{lrrrrrr}
& Intercept & Crude Oil & S\&P500 & Gold & VIX & $\sigma_k$ \\
\toprule
State 1 \\
$\tau= 0.10$ & \textbf{-1.036 (0.280)} & 0.024 (0.021) & \textbf{0.595 (0.096)} & \textbf{0.189 (0.072)} & \textbf{0.029 (0.012)} & 1.433 (0.040) \\
$\tau= 0.50$ & 0.122 (0.158) & 0.031 (0.072) & 0.409 (0.383) & 0.263 (0.249) & 0.009 (0.036) & 1.695 (0.062) \\
$\tau= 0.90$ & \textbf{1.297 (0.061)} & -0.009 (0.020) & \textbf{0.589 (0.088)} & \textbf{0.134 (0.065)} & 0.014 (0.011) & 1.335 (0.041) \\
\midrule
State 2 \\
$\tau= 0.10$ & \textbf{-6.52 (0.060)} & \textbf{-0.256 (0.096)} & \textbf{2.072 (0.476)} & \textbf{1.032 (0.320)} & -0.055 (0.058) & 4.964 (0.157) \\
$\tau= 0.50$ & 0.242 (0.092) & -0.056 (0.055) & \textbf{1.087 (0.357)} & \textbf{0.613 (0.214)} & -0.025 (0.026) & 6.164 (0.169) \\
$\tau= 0.90$ & \textbf{6.244 (0.229)} & 0.017 (0.079) & \textbf{0.948 (0.291)} & \textbf{0.835 (0.249)} & -0.002 (0.041) & 4.692 (0.128) \\
\bottomrule
\end{tabular}%
    }
      \caption{State-specific parameter estimates for three expectile levels, with bootstrapped standard errors (in brackets) obtained over 1000 replications. Point estimates are displayed in boldface when significant at the standard 5\% level.}
       \label{tab:parest}%
\end{table}%

\section{Conclusions}\label{sec:concl} 
The increasing popularity and importance of Bitcoin in the financial landscape have made scholars and practitioners interrogated about its properties and its relation to other assets. In this regard, we contribute to the existing literature in two ways. From a theoretical standpoint, we develop a linear expectile hidden Markov model for the analysis of time series where temporal behaviors of the data are captured via time-dependent coefficients that follow an unobservable discrete homogeneous Markov chain. The proposed method enables us to model the entire conditional distribution of asset returns and, at the same time, to grasp unobserved serial heterogeneity and rapid volatility jumps that would otherwise go undetected. From a practical point of view, we analyze the association between Bitcoin and a collection of global market indices, not only at the average, but also during times of market stress.
\\
Empirically, we find evidence of strong and positive interrelations among Bitcoin returns and S$\&$P500 and Gold, and, at the same time, we observe the capacity of the Bitcoin of working as a weak hedge during high volatility periods, contributing to the existing strands of literature on the subject \citep{baur2018bitcoin, corbet2018exploring, corbet2019cryptocurrencies}. 
Its partial capacity of being a weak hedge but not a safe haven it is consistent with the excess volatility of Bitcoin and indications that assets with no history as a safe haven are unlikely to be considered \qmo safe\qmcsp in an economic or financial crisis \citep{baur2018bitcoin}.
\\
As a possible next step, our methodology could be extended to the hidden semi-Markov model setting where the sojourn-distributions, that is, the distributions of the number of consecutive time points that the chain spends in each state, can be modeled by the researcher using either parametric or non-parametric approaches instead of assuming geometric sojourn densities as in HMMs.
\newpage

\section*{Appendix}
\subsection*{Figures}

\begin{figure}[!h]
\centering
\includegraphics[width=.7\linewidth]{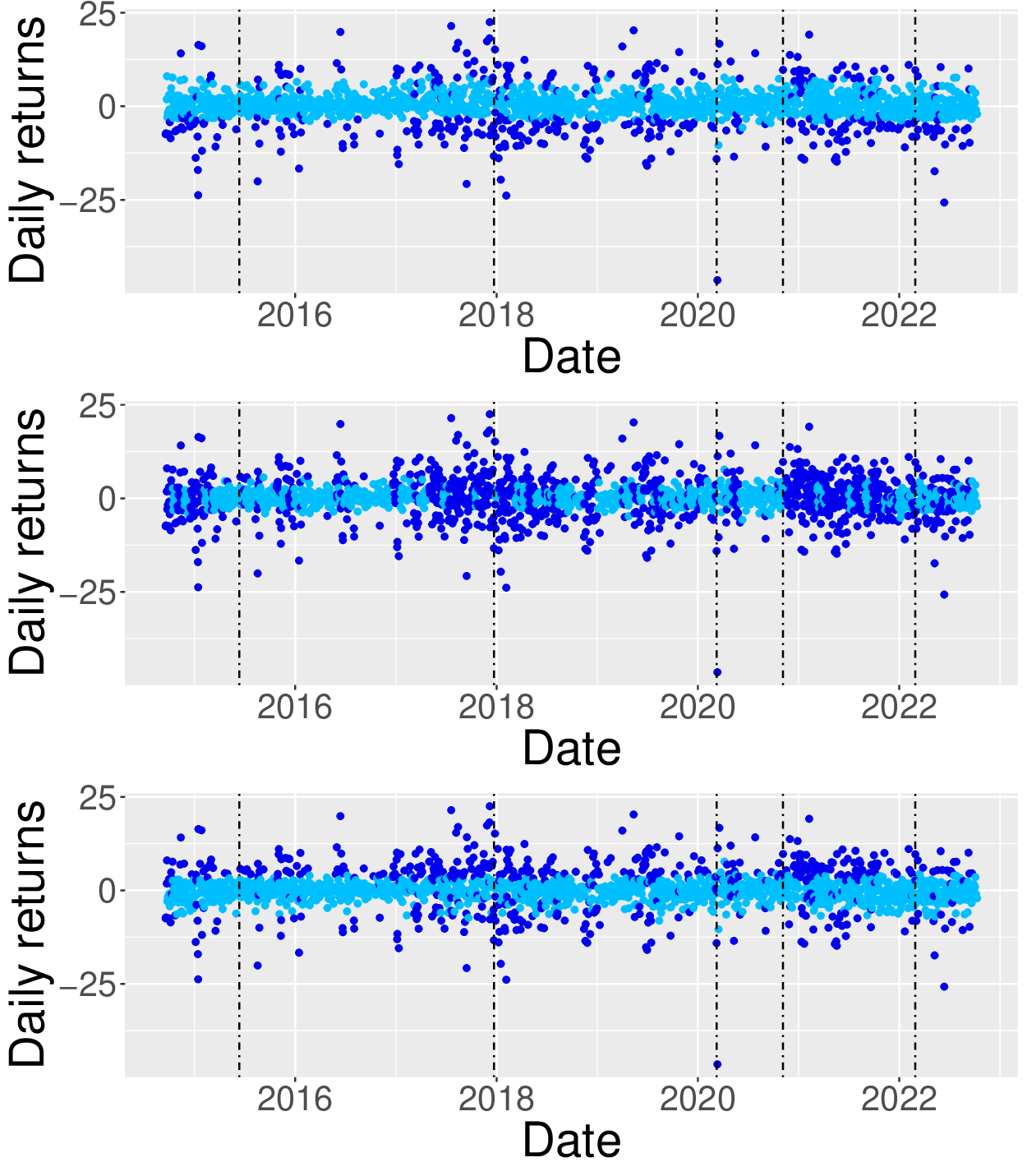}

\caption{From top to bottom, Bitcoin returns series classified according to the Viterbi algorithm for class membership at $\tau = 0.10$, $\tau = 0.50$ and $\tau = 0.90$. Vertical dashed lines indicate globally relevant events in the financial markets that occurred in 2015,06; 2017,12; 2020,03; 2020,11; and 2022,02.}
\label{fig:cl_viterbi}
\end{figure}

\FloatBarrier
\newpage

\bibliographystyle{agsm}
\bibliography{biblio_risk}

\end{document}